\documentclass[
superscriptaddress,
twocolumn,
amsmath,amssymb,
aps,
pra,
]{revtex4-2}

\usepackage{graphicx,framed} 
\usepackage{dcolumn} 
\usepackage{bm} 
\usepackage{braket}
\usepackage{dsfont}
\usepackage{color}
\usepackage{amsthm}
\usepackage{hyperref} 
\usepackage[normalem]{ulem} 
\usepackage{qcircuit}       

\usepackage{amsthm}

\usepackage{comment}

\usepackage{algorithm}
\usepackage[noend]{algpseudocode}
\algdef{SE}[DOWHILE]{Do}{doWhile}{\algorithmicdo}[1]{\algorithmicwhile\ #1}
\makeatletter
\def\algbackskip{\hskip-\ALG@thistlm}
\makeatother

\definecolor{lightblue}{RGB}{73,151,208}
\definecolor{crimson}{RGB}{140,41,53}

\hypersetup{
    colorlinks,
    linkcolor={crimson},
    citecolor={lightblue},
    urlcolor={lightblue}
}

\begin{document}

\preprint{}

\title{Berry Phase in Pathangled Systems}



\author{H. O. Cildiroglu}
\email{cildiroglu@ankara.edu.tr}
\affiliation{Ankara University, Department of Physics Engineering, Ankara, 06100, Turkiye}



\begin{abstract}
We introduce pathangled quantum states, spatially correlated systems governed via production angles, to achieve geometric control of entanglement beyond spin/polarization constraints. By driving the system through cyclic adiabatic evolution of an external parameter in Mach-Zehnder interferometers, we demonstrate that Berry phases and production angles become additional degrees of freedom for Bell correlations. We identify an approximate critical angle $24.97^\circ$ that geometrically manifests the Bell-limit for certain measurement settings, delineating boundaries between local-hidden-variable theories and quantum mechanics. This framework simplifies state preparation while enabling geometry-driven entanglement control, thus providing distinct experimental advantages.
\end{abstract}

\maketitle


The measurement and control of entanglement play a pivotal role in advancing quantum information science and technologies. For two-particle systems, concurrence has been established as a key metric \cite{Wootters1998}, quantifying between 0 (separable states) and 1 (maximally entangled states) to provide a standardized measure for experimental studies \cite{Nielsen2000}. Conventional approaches investigate entanglement in EPR-Bell setups using spin or photon polarizations \cite{Einstein1935, Bell1964, Clauser1969, Clauser1972, Aspect1982, Bertlman2003}. The production-angle-controlled path-entangled states (simply pathangled states) provide a novel framework enabling both theoretical modeling and experimental applications \cite{cildiroglu2025a, Cildiroglu2025b}.

Pathangled systems can be constructed for all kinds of quantum particles (bosons and fermions) using Mach-Zehnder (MZ) type interferometers. In such setups, phase shifters (P) serve as functional analogs to measurements of spin/polarization directions as observables with path-specific detectors \cite{Zeilinger1981, Horne1989}. Angle $\alpha$ directly modulates spatially correlated wavefunctions, allowing for precise experimental control over quantum concurrence with broad applications from particle physics to quantum gravity searches.

Here, we investigate Bell inequalities (BI) within pathangled systems by incorporating the Berry phase ($\gamma$) in single/double beam splitter MZ-type setups \cite{Berry1984}. This framework establishes $\alpha$ and $\gamma$ as new degrees of freedom to control the Bell function ($S$) under certain measurement settings, offering rigorous characterization via their dependencies. We identify a critical production angle $\alpha_c\approx 24.97^o$, the system reaches $S=2$, signifying a geometric analog to the Bell-limit between local hidden-variable models (LHVM) and quantum mechanics (QM). Thereby, for $\alpha_c<\alpha < \pi/2-\alpha_c$, the system can be operated with nonlocal QM properties beyond classical limits.

The letter organized as follows: First, we introduce pathangled states and geometric phases under adiabatic evolution. The $S$ parameter is then derived for single and double-BS MZ scenarios, testing BI under fixed measurement settings. Finally, we present the $\alpha_c$ as a geometric analog to the Bell parameter and its geometric demarcation of quantum-classical boundaries.

In two-dimensions, spatial correlations dictate that a quantum mechanical particle (quanton) emitted toward the upper left at an arbitrary $\alpha$ direction ($|u_\alpha\rangle_L$) must be accompanied by its pair toward the lower right ($|d_\alpha\rangle_R$). Conversely, a downward-left quanton ($|d_\alpha\rangle_L$) corresponds to an upward-right counterpart ($|u_\alpha\rangle_R$). Hence, the normalized pathangled states  $|\psi\rangle = |u_\alpha\rangle_R \otimes |d_\alpha\rangle_L + |d_\alpha\rangle_R \otimes |u_\alpha\rangle_L$ can be written as,

\begin{align}
\label{1}
\ket{\psi} = \sqrt{\frac{1 - C(\alpha)}{2}} |\chi^+\rangle + \sqrt{\frac{1 + C(\alpha)}{2}} |\varphi^+\rangle
\end{align}

\noindent where $(1\hspace{2mm} 0)^T \equiv \ket{0}$ and $ (0\hspace{2mm} 1)^T \equiv \ket{1}$;  $|\chi^\pm\rangle= \frac{1}{\sqrt{2}} \left(|00\rangle \pm |11\rangle\right)$ and $|\varphi^\pm\rangle = \frac{1}{\sqrt{2}} \left(|01\rangle \pm |10\rangle\right)$ are the Bell states, and $C(\alpha) = \frac{1-\cos^2(2\alpha)}{1 + \cos^2(2\alpha)}$ is the entanglement concurrence in the basis aligned along the BS plane,

\begin{equation}
\label{2}
|u_\alpha\rangle= \begin{pmatrix} \cos\left(\frac{\pi}{4} - \alpha\right) \\ \sin\left(\frac{\pi}{4} - \alpha\right) \end{pmatrix}, \hspace{2mm} 
|d_\alpha\rangle= \begin{pmatrix} \sin\left(\frac{\pi}{4} - \alpha\right) \\ \cos\left(\frac{\pi}{4} - \alpha\right) \end{pmatrix}
\end{equation}

\noindent In this representation, $\alpha = 0$ yields $C = 0$, corresponding to a product state, while $\alpha = \frac{\pi}{4}$ leads to $C = 1$, indicating a maximally entangled state. For $\alpha=\pm \frac{\pi}{4}$, quantons are aligned along the inputs of BS ($\ket{0}$, $\ket{1}$), whereas for $\alpha=0$ and $\frac{\pi}{2}$, the particles emerge in horizontal and vertical directions respectively that corresponds to the case where all quantons are sent in the same direction. The detectors are aligned with the BS output ports ($\ket{D_0} \equiv (1,0)^T$, $\ket{D_1} \equiv (0,1)^T$), oriented at $\frac{\pi}{4}$ to the horizontal (see for detailed analysis ref: \cite{cildiroglu2025a, Cildiroglu2025b}).

\begin{figure*}[ht]
    \centering \includegraphics[width=\textwidth]{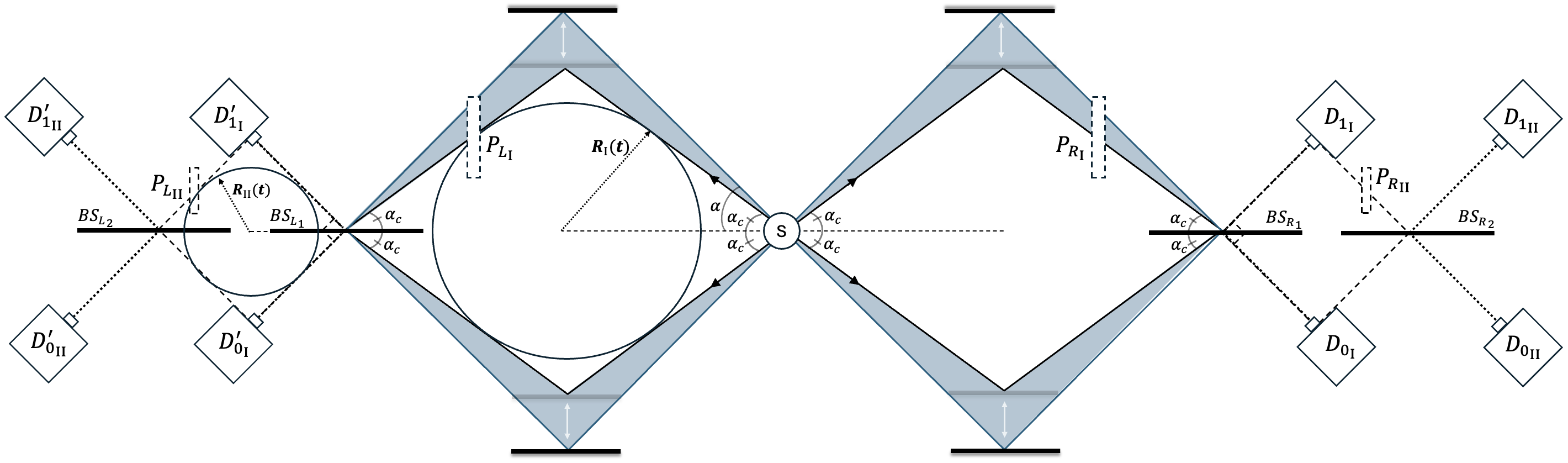}
    \caption{Schematic representation of pathangled systems. In the first scenario, particles adiabatically driven by $R_1(t)$ acquire a geometric phase and are detected after $P_{LR_1}$ and $BS_1$. In the second scenario, cyclic evolution under $R_2(t)$ with $P_{LR_2}$ enables geometric/topological phase observation in closed MZ trajectories. Bell limit occurs at production angle $(\alpha)$ equals $\alpha_c\approx24.97^\circ$, quantum nonlocality manifests for $\alpha_c <\alpha<\pi/2-\alpha_c$ in blue regions with certain retarder settings.}
    \label{Fig1}
\end{figure*}

On the other hand, geometric phases, which arise from cyclic adiabatic quantum evolution, are fundamentally linked to the holonomy of the state's line bundle via the integral of its connection over the parameter space, and represent a profound and broadly significant phenomenon in physics. Analyzing how these phases affect pathangled systems could provide novel insights into quantum foundations with potential applications that bridge particle physics with the quest for quantum gravity. Accordingly, BI offer a rigorous experimental framework for characterizing these phase-dependent quantum correlations, enabling precise verification of their nonclassical nature.

\newpage

In an EPR-Bohm type setup, pathangled particles originating from the same source, wlog those propagating leftward (see Fig. \ref{Fig1}), undergo adiabatic cyclic evolution driven by an external parameter R(t), during which each eigenstate acquires a phase factor comprising geometric ($\gamma_\pm$) and dynamical ($\theta_\pm$) components.

\begin{align} 
\label{3}
\ket{0}_{t=0} \rightarrow \ket{0}_{t=\tau} = e^{i\gamma_+} e^{i\theta_+} \ket{0}_{t=0} \nonumber\\
\ket{1}_{t=0} \rightarrow \ket{1}_{t=\tau} = e^{i\gamma_-} e^{i\theta_-} \ket{1}_{t=0}
\end{align}

\noindent Here $\gamma_-=-\gamma_+ - 2\pi$ and $\theta_-=-\theta_+=\frac{1}{\hbar}E_\pm \tau$ and the dynamical phase typically dominate the geometric phase cancels out as it reduces to a global phase factor via symmetric trajectory design \cite{Bertlman2003}. For the parameter's two half periods of rotation,

\begin{equation}
\label{4}
\ket{0} \rightarrow e^{i\gamma_+} \ket{0}, \hspace{3mm} \ket{1} \rightarrow e^{i\gamma_-} \ket{1}
\end{equation}

\noindent the Bell states in \eqref{1} are expressed with a common phase factor $e^{i\gamma_-}$, where $\gamma_+ \equiv \gamma$ can be defined by convention.

\begin{align}
\label{5}
    \ket{\chi^+}_\tau &\rightarrow \frac{1}{\sqrt{2}}\left[e^{2i\gamma}\ket{00}+\ket{11}\right] \nonumber\\
    \ket{\varphi^+}_\tau &\rightarrow \frac{1}{\sqrt{2}} \left[e^{2i\gamma}\ket{01}+\ket{10}\right]
\end{align}

We now consider MZ setups in two scenarios to test the BI. Within this framework, quantons are detected after being directed into single/double BS configurations. The operations of symmetric and lossless BSs with two input and two output ports, as well as phase-shifting elements 
P, which are positioned in each interferometer on both sides, are described by unitary matrix representations \cite{Zeilinger1981}.

\begin{equation}
\label{6}
BS=\frac{1}{\sqrt{2}}\left(\begin{array}{cc} 1 & i\\
i & 1 \end{array}\right), \hspace{3mm}
P=\left(\begin{array}{cc} e^{i\vartheta_{L(R)}} & 0 \\
0 & 1 \end{array}\right)
\end{equation}

\noindent \textbf{{Single-BS systems:}} In the first scenario, the pathangled quantons traverse $P_{LR_1}$ and arrive at the first BSs in the state $|\psi_{\rm I}\rangle = (BS_{R_1} \otimes BS_{L_1})(P_{R_{\rm I}} \otimes P_{L_{\rm I}})|\psi\rangle$, acquiring a geometric Berry phase during adiabatic cyclic evolution driven by the external parameter $R_{\rm I}(t)$ (or its planar projection in higher dimensions) along symmetric paths in the left MZ plane. Defining, $\vartheta_\pm=\vartheta_R \pm \vartheta_L$ and $\kappa_\pm=\frac{\vartheta_\pm}{2} \pm \gamma_\pm$ the state becomes at $t=\tau$,

\begin{align}
\label{7}
\left| \psi_{\rm I}^\tau \right\rangle = \sqrt{\frac{1 - C}{2}} \left[ \cos{\kappa_+}\left| \varphi^{+} \right\rangle + \sin\kappa_+\left| \chi^{-} \right\rangle \right] \\+ \sqrt{\frac{1 + C}{2}}\left[ \sin\kappa_-\left| \varphi^{-} \right\rangle + \cos\kappa_-\left| \chi^{+} \right\rangle \right] \nonumber
\end{align}

\noindent After eliminating the common phase factor, the joint-detection amplitudes are given by $A_{\rm I}(D_{i}, D^{'}_{j})=\langle D_{i_{\rm I}} D^{'}_{j_{\rm I}}|\psi_{out}\rangle$ with $i,j=0,1$,

\begin{align}
    \label{8}
    A_{\rm I}(D_{0(1)}, D^{'}_{0(1)})= (-)\frac{\sqrt{1-C}}{2} \sin{\kappa_+}+ \frac{\sqrt{1+C}}{2} \cos{\kappa_-} \nonumber\\
    A_{\rm I}(D_{0(1)}, D^{'}_{1(0)})= \frac{\sqrt{1-C}}{2} \cos{\kappa_+}\stackrel{(-)}{+} \frac{\sqrt{1+C}}{2} \sin{\kappa_-}
\end{align}

\noindent Thus, using these amplitudes, the joint-detection probabilities, $P_{\rm I}(D_{i}, D^{'}_{j})= \left| A_{\rm I}(D_{i}, D^{'}_{j}) \right|^2$ are obtained as:

\begin{align}
\label{9} 
P_{\rm I}(D_{0(1)},D_{0(1)}')=&\frac{1}{4}\Bigg[1-\bigg(\frac{1-C}{2}\bigg)\cos{2\kappa_+}+\bigg(\frac{1+C}{2}\bigg)\cos{2\kappa_-}\bigg] \nonumber \\ &\stackrel{(-)}{+}\frac{\sqrt{1-C^2}}{2} \sin{\kappa_+} \cos{\kappa_-} \nonumber \\
P_{\rm I}(D_{0(1)},D_{1(0)}')=&\frac{1}{4}\Bigg[1+\bigg(\frac{1-C}{2}\bigg)\cos{2\kappa_+}-\bigg(\frac{1+C}{2}\bigg)\cos{2\kappa_-}\Bigg]\nonumber\\ &\stackrel{(-)}{+}\frac{\sqrt{1-C^2}}{2} \sin{\kappa_-} \cos{\kappa_+} 
\end{align}

\noindent For $C=1$ (or equivalently for $\alpha=\pi/4$) the detection probabilities reduce to $P_{\rm I}(D_{0(1)},D_{0(1)}')=\frac{1}{4}\left[1+\cos{2\kappa_-}\right]$ and $P_{\rm I}(D_{0(1)},D_{1(0)}')=\frac{1}{4}\left[1-\cos{2\kappa_-}\right]$. When $\kappa_-=\kappa_+=0$, this yields $P_{\rm I}(D_{0(1)},D_{0(1)}')=1/2$ and $P_{\rm I}(D_{0(1)},D_{1(0)}')=0$ indicating that the system is maximally entangled. The expectation values for joint measurements are therefore $E_{\rm I}(\vartheta_L, \vartheta_R) = \sum_{i,j=0,1} (-1)^{i+j} P_{\rm I}({D_i}^{\vartheta_L}, {D'_j}^{\vartheta_R})$,

\begin{align}
\label{10}
E_{\rm I} =-\bigg(\frac{1-C}{2}\bigg)\cos{2\kappa_+} +\bigg(\frac{1+C}{2}\bigg) \cos{2\kappa_-}
\end{align}

\noindent To experimentally isolate the geometric phase effects on the system, one can set all phases to zero ($\vartheta_{L(R)}=0$). In this case, $\kappa_\pm=\gamma$, and the equations \eqref{9} and \eqref{10} simplify to $P(D_i, D'_j) = \frac{1}{4} \left[1 + (-1)^{i+j} C \cos{2\gamma} + (-1)^{i} \sqrt{1 - C^2} \sin{2\gamma}\right]$ and $E = C\cos{2\gamma}$ respectively. The correlation function 
S for the BI tests is defined as:

\begin{align}
\label{11}
S(\vartheta_L,\vartheta_R,\vartheta_L',\vartheta_R';\alpha,\gamma)=&|E(\vartheta_L,\vartheta_R)-E(\vartheta_L,\vartheta_R')|\nonumber\\+&|E(\vartheta_L',\vartheta_R)+E(\vartheta_L',\vartheta_R')|
\end{align}

\noindent This exhibits maximal violation of the BI for retarder settings $(\vartheta_L,\vartheta_R,\vartheta_L',\vartheta_R')=(0, \frac{\pi}{4},\frac{\pi}{2}, \frac{3\pi}{4})$, which are analogous to optimal spin/polarization measurement directions in quantum systems,

\begin{equation}
\label{12}
S_{\rm I}(\alpha,\gamma)=\sqrt{2}+C\sqrt{2}\thinspace \left|\cos{2\gamma}\right|
\end{equation}

\begin{figure*}[ht]
    \centering \includegraphics[width=0.95\textwidth]{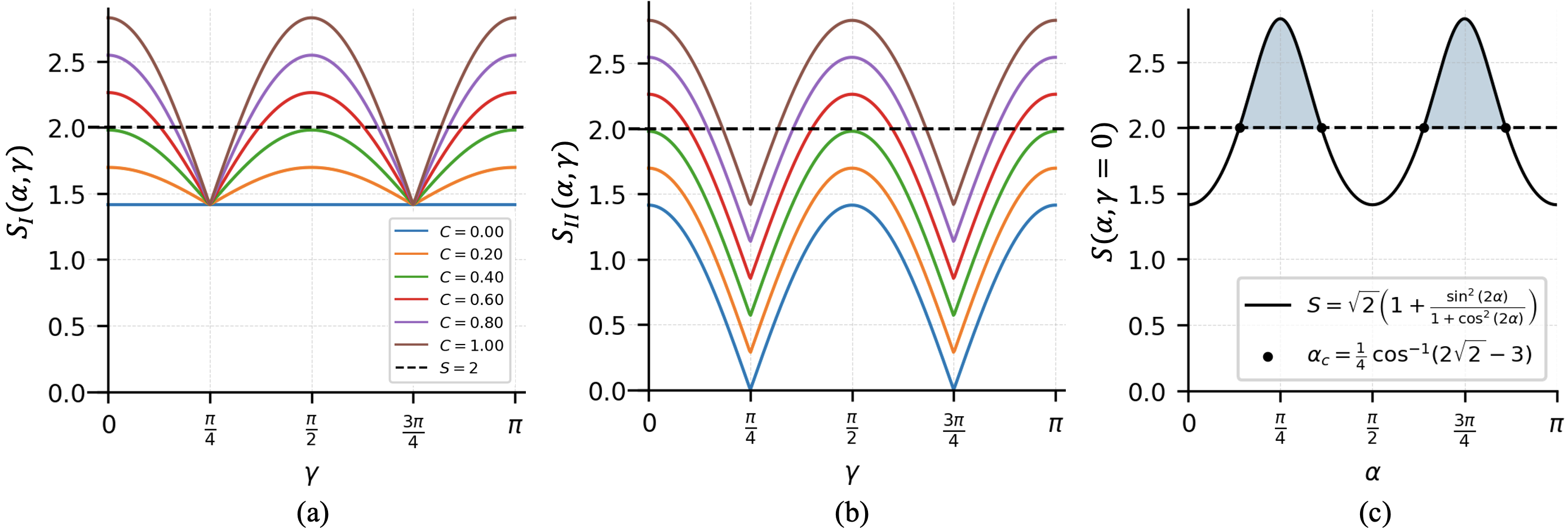}
    \caption{Correlation function $S(\alpha,\gamma)$ under distinct scenarios. (Left) Plots of $S_{\rm I}(\alpha,\gamma)= \sqrt{2}+\sqrt{2}C(\alpha)\left|\cos{2\gamma}\right|$ vs $\gamma$ for fixed $C(\alpha)$ values. The shared control of $C(\alpha)$ and $\gamma$ over the same term bounds the minimum of $S$ at $\sqrt{2}$. (Center) Plots of $S_{\rm II}(\alpha,\gamma)=C\sqrt{2}+\sqrt{2}\left|\cos{2\gamma}\right|$ vs $\gamma$ for fixed $C(\alpha)$, with colors matching the left panel.  Independent control over $C(\alpha)$ and $\gamma$ enables access to all correlation parameters in the range $(0,2\sqrt{2})$. (Right) Both scenarios attain the Bell-limit $(S=2)$ at $C\approx 0.4$ (corresponding critical angle $\alpha_c \approx 24.97^\circ$). For $\alpha<\alpha_c$, the Bell-limit $S=2$ cannot be exceeded regardless of $\gamma$. For $\gamma=0$, the blue region here is consistent with that in Fig. ~\ref{Fig1}, exhibiting identical quantum correlations. Within $\alpha_c < \alpha < \pi/2 - \alpha_c$, particles exhibit nonlocal QM behaviors. Thus, $\alpha_c$ constitutes an operational analog to certain retarder configurations for $S$. Accordingly, at $\alpha=\pi/4$, the Tsirelson bound $S=2\sqrt{2}$ is saturated.} 
    \label{Fig2}
\end{figure*}

\noindent The expression explicitly reveals the influence of the Berry phase on pathangled systems, demonstrating that the BI violations can be controlled via both $C(\alpha)$ and $\gamma$ (see Fig.~\ref{Fig2}-left). For maximally entangled states $(C=1)$, it reduces to $S=2\sqrt{2}\cos{2\gamma}$. This enables direct geometric phase control in any quanton system by offering enhanced experimental feasibility, with a spanning $\left( 2,2\sqrt{2}\right)$ as $\gamma$ is varied, demonstrating quantum contextual behavior.

\vspace{2mm}

\noindent \textbf{Double-BS systems:}  In the second scenario, we now investigate pathangled quantons in double BS - MZ interferometers. The initial state \eqref{1} evolves under the transformation $|\psi_{\rm II}\rangle=(BS_{R_2}\otimes BS_{L_2})(P_{R_{\rm II}} \otimes P_{L_{\rm II}})(BS_{R_1} \otimes BS_{L_1})|\psi\rangle$. During adiabatic evolution driven by $R_{\rm II}(t)$ ($R_{\rm I}=0)$, the system acquires a geometric phase,

\begin{align}
\label{13}
| \psi&_{\rm II}^\tau \rangle = \sqrt{\frac{1 - C}{2}}
\bigg[\begin{matrix}
\cos{\gamma}\left[\cos{\frac{\vartheta_-}{2}} \ket{\chi^+}+\sin{\frac{\vartheta_-}{2}} \ket{\varphi^-}\right] \nonumber
\\-i\sin{\gamma}\left[\cos{\frac{\vartheta_+}{2}} \ket{\chi^-}-\sin{\frac{\vartheta_+}{2}} \ket{\varphi^+}\right]
\end{matrix}\bigg]\\
+&\sqrt{\frac{1 + C}{2}}\bigg[
\begin{matrix}
\cos{\gamma}\left[\sin{\frac{\vartheta_+}{2}} \ket{\chi^-}+\cos{\frac{\vartheta_+}{2}} \ket{\varphi^+}\right]
\\+i\sin{\gamma}\left[\sin{\frac{\vartheta_-}{2}} \ket{\chi^+}-\cos{\frac{\vartheta_-}{2}} \ket{\varphi^-}\right]
\end{matrix}\bigg]
\end{align}

\noindent Thus, the joint-detection amplitudes, 

\begin{align}
    \label{14}
    A_{\rm II}(D_{0(1)}, &D^{'}_{0(1)})= \frac{\sqrt{1-C}}{2} \bigg[ \cos{\gamma}\cos{\frac{\theta_-}{2}}\stackrel{(+)}{-}i\sin{\gamma}\cos{\frac{\theta_+}{2}}\bigg] \nonumber
    \\+&\frac{\sqrt{1+C}}{2} \bigg[(-)\cos{\gamma}\sin{\frac{\theta_+}{2} }+i\sin{\gamma}\sin{\frac{\theta_-}{2}}\bigg]\nonumber
    \\
A_{\rm II}(D_{0(1)}, &D^{'}_{1(0)})= \frac{\sqrt{1-C}}{2} \bigg[(-) \cos{\gamma}\sin{\frac{\theta_-}{2}}+i\sin{\gamma}\sin{\frac{\theta_+}{2}}\bigg] \nonumber
    \\+&\frac{\sqrt{1+C}}{2} \bigg[\cos{\gamma}\cos{\frac{\theta_+}{2} }\stackrel{(+)}{-}i\sin{\gamma}\cos{\frac{\theta_-}{2}}\bigg]
\end{align}

\noindent and the corresponding probabilities are given as;

\begin{align}
    \label{15}
    P_{\rm II}(D_{0(1)}, &D^{'}_{0(1)}) = \frac{1}{4}\bigg[1 - C\cos{\vartheta_L}\cos{\vartheta_R} + \sin{\vartheta_L}\sin{\vartheta_R}\cos{2\gamma} \nonumber\\&\stackrel{(-)}{+}\sqrt{1 - C^2}\big(\sin{\vartheta_{R}}+\sin{\vartheta_{L}} \cos{2\gamma} \big)\bigg]\nonumber \\
    P_{\rm II}(D_{0(1)}, &D^{'}_{1(0)}) = \frac{1}{4}\bigg[1 + C\cos{\vartheta_L}\cos{\vartheta_R} - \sin{\vartheta_L}\sin{\vartheta_R}\cos{2\gamma} \nonumber\\&\stackrel{(+)}{-}\sqrt{1 - C^2}\big(\sin{\vartheta_{R}}-\sin{\vartheta_{L}} \cos{2\gamma} \big)\bigg]
\end{align}

\noindent Accordingly, the expectation values are derived in terms of $C$ and $\gamma$:

\begin{align}
\label{16}
    E_{\rm II}=-C\cos{\vartheta_L}\cos{\vartheta_R}+\sin{\vartheta_L}\sin{\vartheta_R}\cos{2\gamma}
\end{align}

\noindent Finally, for certain phase settings $(\vartheta_L,\vartheta_R,\vartheta_L',\vartheta_R')=(0, \frac{\pi}{4},\frac{\pi}{2}, \frac{3\pi}{4})$, equation \eqref{11} takes the form:

\begin{equation}
\label{17}
S_{\rm II}(\alpha,\gamma)=C\sqrt{2}+\sqrt{2}\left|\cos{2\gamma}\right| 
\end{equation}

The findings highlight the distinct roles of $C(\alpha)$ and $\gamma$ in determining the Bell parameter, where both terms contribute separate $\sqrt{2}$ factors to the correlation function. It provides more flexible control compared to path-spin entanglement implementations \cite{Bertlman2003}, where control is constrained by fixed spin-path couplings and limited parameter tuning. For $C=0$, $S=\sqrt{2}\left|\cos{2\gamma} \right|$ remains geometric phase-dependent that contrasts with entanglement-dependent effects that vanish for product states. Similarly, when $\gamma=(2n+1)\pi/4$ ($n\in\mathbb{Z}$), it becomes $S=C\sqrt{2}$. The maximal violation of BI $S=2\sqrt{2}$ is achieved when $C=1$ and $\gamma=n\pi/2$ (see Fig.~\ref{Fig2}-center). While the previous setup—analogous to single-slit diffraction—features open trajectories restricting topological phases to an unobservable global phase, this configuration operates as a double-slit interferometer for individual quantons. Its closed trajectories enable direct observation of both geometric and topological phases, while reproducing established experimental results \cite{Bertlman2003, Hasegawa2012, Cildiroglu2021, Cildiroglu2024} under appropriate $C(\alpha)$ conditions. This provides a complete analogy to spin/polarization entanglement experiments with a new degree of freedom.

\textbf{Discussion:} We present a novel framework for analyzing pathangled systems through the integration of geometric phase into the concurrence-based description of two-quanton systems. Unlike conventional approaches that rely on spin/polarization degrees of freedom, our formulation leverages spatial correlations defined through production angles, demonstrating entanglement modulation via purely geometric parameters. We note that the Berry phase does not affect the amount of entanglement or nonlocality that quantified by maximal BI violations, but instead provides a mechanism to leverage production angles as a novel degree of freedom to actively control these properties.

Here, we compute the correlation functions from joint-detection probabilities in single/double BS scenarios. The resulting expressions \eqref{12} and \eqref{17} depend on $C$ and $\gamma$ exhibit distinct functional forms stemming from their trajectory topologies despite comparable tunability for BI violations. To illustrate, consider a geometric phase generated the interaction Hamiltonian $\Delta H = \frac{\mu}{2} B(t) \sigma$ under an adiabatically varying magnetic field ($R_{1(2)}(t)=B(t)$). In both scenarios, the joint-spin measurements along arbitrary directions $\vartheta_{L(R)}$ 
relative to the eigenbasis of the interaction term retain $\gamma$, enabling explicit control of $S$. Symmetric trajectories inherently eliminate dominant dynamical phases, precluding requirements for echo-based methods. Furthermore, in systems \cite{Aharonov1959, Aharonov1984} with magnetic dipoles moving around an infinitely long and thin linear charge distribution $\lambda$ ($R_{\rm II}(t)=E$), the topological phase contribution $(\delta=\mu\lambda)$ from the interaction $\Delta H = \frac{\mu}{2} \sigma_{\mu \nu}F^{\mu\nu}$ reduces to a global phase factor in the open trajectories of the single-BS configuration and vanishes in \eqref{10} and \eqref{11}.  However, in the second scenario, the phase yields observable effects and takes the value $\gamma = \mu\lambda$, directly controlling \eqref{16} and \eqref{17}. For $C(\alpha=\frac{\pi}{4})=1$, our results align with geometric and topological phase signatures in entangled spin systems \cite{Bertlman2003, Cildiroglu2021, Cildiroglu2024}, as recently verified experimentally in \cite{Kumar2025}.

Our results reveal that when the geometric phase vanishes $(\gamma=0)$, the concurrence $C(\alpha)=\frac{1-\cos^2{2\alpha}}{1+\cos^2{2\alpha}}$, causing both \eqref{12} and \eqref{17} to converge at $S=\sqrt{2}(1+C(\alpha))$. This establishes a critical production angle $\alpha=\frac{1}{4}\cos^{-1}{(2\sqrt{2}-3)}\approx 24.97^\circ$ for pathangled states, where $S=2$ precisely demarcates from the LHVM and QM boundaries. Consequently, BI are violated for $\alpha>\alpha_c$: For quantons confined to the blue regions in Fig ~\ref{Fig1} , $S$ attains values exclusively in the blue violation zones in Fig ~\ref{Fig2}-right, and LHVM compliance requires $\alpha\leq \alpha_c$. Thus, 
$\alpha_c$ serves as the equivalent analog of the Bell function in pathangled systems.

In conclusion, pathangled systems function universally across over all quantons (fermion, boson). This framework offers practical advantages over spin-correlated systems. Spatial correlations enable simpler state preparation and more direct control through production angles. The pathangled states in single/double BS systems enable diverse implementations of quantum geometric and topological phase phenomena. In turn, the phases enter correlation functions as an additional degree of freedom, providing experimental flexibility with production angles. The critical angle provides a geometric criterion for QM nonlocality, functionally replacing the parameter $S$ where $\alpha > \alpha_c$ directly indicates quantum violations under specific measurement settings. These interferometric configurations may facilitate future work across the bridge between high and low energy studies from particle physics to searching for quantum gravity in the sky.







\bibliography{refs}

\end{document}